\documentclass[pdflatex,sn-mathphys-num]{sn-jnl}

\usepackage{graphicx}%
\usepackage{multirow}%
\usepackage{amsmath,amssymb,amsfonts}%
\usepackage{amsthm}%
\usepackage{mathrsfs}%
\usepackage[title]{appendix}%
\usepackage{xcolor}%
\usepackage{textcomp}%
\usepackage{manyfoot}%
\usepackage{booktabs}%
\usepackage{algorithm}%
\usepackage{algorithmicx}%
\usepackage{algpseudocode}%
\usepackage{listings}%
\usepackage{enumitem}%
\usepackage{nicefrac}%
\usepackage{physics}%
\usepackage{comment}%

\theoremstyle{thmstyleone}%

\theoremstyle{thmstyletwo}%

\theoremstyle{thmstylethree}%

\begin{document}

\title{A Quantum--Analogue Formalism for Modeling Supraliminal Information Processing}


\author*[1]{\fnm{Vasily} \sur{Lubashevskiy}}\email{vlubashe@tiu.ac.jp}
\author[2]{\fnm{Ihor} \sur{Lubashevsky}}\email{ilubashevskii@hse.ru}

\equalcont{The authors contributed equally to this work.}

\affil[1]{\orgdiv{Institute for International Strategy}, \orgname{Tokyo International University}, \orgaddress{\street{4 Chome-42-31 Higashiikebukuro, Toshima}, \city{Tokyo}, \postcode{170-0013}, \country{Japan}}}

\affil[2]{\orgdiv{Tikhonov Moscow Institute of Electronics and Mathematics}, \orgname{HSE University}, \orgaddress{\street{34 Tallinskaya str.}, \city{Moscow}, \postcode{123458}, \state{State}, \country{Russia}}}

\abstract{We develop a novel cloud-function formalism describing the dynamical relationship between sensory-information processing in large-scale brain networks (supraliminal processing) and the content of the mental representation of an observed object. The formalism combines elements of neural field theory for large-scale neural activity with the spatial characteristics of perceived objects and their embedding in the environment from the first-person perspective.
The cloud function is characterized by two key features: (\textit{i}) its spatial structure inherits properties of the perceived physical object, and (\textit{ii}) its temporal evolution is governed by regularities reflecting intrinsic properties of large-scale neural activity.
The governing equation for the cloud function is based on a neural-field model with polynomial nonlinearities and global phase-shift invariance of neural-pattern oscillations. Its structure may be interpreted as a Schr\"odinger-type equation with a nonlinear non-Hermitian Hamiltonian supplemented by terms analogous to those of the Lotka--Volterra model.
The proposed approach is applied to the change-of-mind phenomenon in decision-making, in which an initial choice may be revised during its execution. Changes of mind are explained as arising from the interplay between fast preconscious sensory processing and slower conscious comparison of alternatives, consistent with neurophysiological evidence for continuous post-decisional evidence accumulation. The necessity of incorporating cloud-function self-interaction is also discussed.
}




\maketitle

\section{Introduction}\label{sec1}

Over recent years, substantial progress has been made in elucidating the spatio--temporal correspondences between patterns of neural activity and basic features of phenomenal consciousness \citep[][for a review]{Northoff2024}. In particular, the novel \textit{theory of common currency} \citep{Northoff2025b} posits that neural and mental phenomena share intrinsic spatio--temporal features which provide a direct basis for their correspondence, without recourse to any intermediate, non-temporal or non-spatial explanatory level. Put differently, the brain and conscious experience are proposed to possess intrinsic spatio--temporal frameworks---distinct from physical spacetime---that share fundamental dynamical characteristics, such as scale-free temporal hierarchies, fluctuations, and synchrony. These commonalities are argued to provide a \textit{dynamical bridge} between first-person and third-person perspectives \citep{Northoff2025Gap}, to constitute the basis of the spatio--temporal theory of consciousness \citep{Northoff_2022}, and to be grounded in the active inference approach to describing this neural--mental correspondence \citep{Lutz2024}.

Shepard's second-order isomorphism of internal representations \citep{Shepard_1970} may be viewed as a precursor to contemporary notions of dynamical psycho-neural correspondence. Without explicitly distinguishing the internal spatio-temporal structure of mental images from that of physical reality, this framework posits that temporal variations in observed objects (e.g., position or shape) and their corresponding internal representations admit a shared \textit{parameterization} \citep{Edelman_1998}. In contrast, the concept of Neural Correlates of Consciousness (NCC) \citep{Crick1990,Crick_1998}, particularly in its content-specific formulation, emphasizes the entities involved in this correspondence by relating patterns of neural activity---arising from sensory processing---to observed objects \citep[e.g.,][for a review]{Almeida_2022}.

Recently, we proposed a novel approach to characterizing this dynamical bridge between neural activity patterns and subjective experience from the first-person perspective \citep{LubashNatPhen2025}. The central idea is to use properties of mental images to model the neural processing of sensory information, with particular emphasis on its final stage, which we term \textit{supraliminal information processing}. Following \citet{Kouider_2007}, we use \textit{supraliminal} to denote the union of preconscious (supraliminal and unattended) and conscious (supraliminal and attended) stages within the subliminal--preconscious--conscious taxonomy \citep{Dehaene2006}.

The noted role of attention makes it necessary to emphasize that top-down selective attention and consciousness, although intertwined within supraliminal information processing, constitute distinct phenomena with different functions and neural mechanisms \citep[][for a review]{Tsuchiya2016}. From the perspective of dynamical systems, supraliminal processing of sensory information may thus be conceived as an integrated process in which top-down selective attention modulates the effective potential landscape, whose structure is shaped by early stages of sensory processing \citep[e.g.,][for a detailed discussion]{LubashNatPhen2025}. Within this framework, information processing attains the level of conscious access when the underlying neural dynamics reaches stable or metastable attractor states \citep[][for a review]{Esteban2026}. Accordingly, when the level of attention is sufficient to enable conscious access (as discussed in more detail in Sec.~\ref{sec2}), preconscious processing can be interpreted as a transient phase of the same dynamical evolution---precisely the case considered here.

Our approach aims to describe the evolution of mental images arising from the sensory perception of external objects. However, such a description cannot be confined to elements analyzed solely from the first-person perspective because the intrinsic space and time of these images---space--time clouds---lack point-like elements, such as spatial points or temporal instants. This absence reflects the inherent uncertainty of human perception, which is grounded in the neural mechanisms of sensory information processing \citep{LubPlav2021}.

The absence of point-like instants precludes the direct application of dynamical-systems formalism to mental images considered in isolation. To overcome this limitation, we adopt the third-person perspective and consider high-level neural representations evolving during supraliminal sensory processing. The neural--mental correspondence prompts us to couple these representations with the spatial projections of the corresponding mental images, which we term \textit{cloud functions} \citep{LubashNatPhen2025}. The dynamics of cloud functions should admit joint parameterization with those of the corresponding neural representations. Accordingly, a cloud function (\textit{i}) inherits the spatial structure of the underlying space--time cloud and (\textit{ii}) becomes defined at each instant of time during supraliminal information processing, thereby opening the way to the use of dynamical-systems formalism, including field theory.

The absence of dimensionless spatial points leads us to treat cloud functions as entities with internal integrity. This motivates the use of a Hilbert space as a formal framework in which cloud functions serve as basic elements, rendering the proposed formalism formally analogous, in key respects, to quantum mechanics.
   
The purpose of the present paper is to elucidate the formalism of cloud functions in detail and to demonstrate that the resulting dynamics can be cast in the form of a nonlinear, dissipative Schr\"odinger equation, thereby establishing a principled link between supraliminal neural processing and a well-studied class of dynamical equations. In addition, the proposed formalism is illustrated through the modeling of \textit{changes of mind}, as observed in binary categorization experiments in which subjects revise previously made decisions during their execution \citep{Resulaj2009,Fleming2016chmind}.

\section{Background}\label{sec2}

In this section, we briefly outline the premises underlying our mathematical constructions, which we have not examined previously \citep{LubashNatPhen2025}.

\textbf{1}. Recurrent Processing Theory \citep{Lamme2006,Lamme2010} links consciousness to interactions between feedforward (bottom-up) and feedback (top-down) processes and distinguishes four stages of visual processing. During the third stage, information is integrated across multiple visual areas, giving rise to phenomenal consciousness, whereas the fourth stage involves ignition and global broadcasting, associated with access consciousness. Accordingly, high-level representations formed at these stages may share a similar multiscale structure within the brain's intrinsic space. Within Global Neuronal Workspace Theory \citep{Dehaene2011,Mashour2020}, this view is supported by the role of top-down attentional amplification, which enables ignition and global broadcasting \citep[e.g.,][for a review]{Mudrik2025}.

Taking these features into account, we extend the concept of cloud functions from the third (preconscious) to the fourth stage of information processing. Much of the transformation at the latter stage remains inaccessible to  awareness and unfolds unconsciously \citep{Sklar2021}. Determining which aspects of cloud-function dynamics enter phenomenal or access consciousness requires separate analysis. Here, as previously \citep{LubashNatPhen2025}, we assume that conscious content is represented by the cloud function when its dynamics approaches a steady-state manifold.

\textbf{2}. The novel formalism of connectome harmonics \citep{Atasoy2016,Atasoy2017} describes spatio--temporal brain dynamics in terms of Laplacian eigenmodes of structural connectivity \citep[][for a review]{Atasoy2018Rev,Luppi2023}. Low-frequency oscillations, e.g., the delta, theta, and possibly alpha bands, are associated with modes of small eigenvalues, reflecting large-scale activity patterns and long-range correlations, often linked to conscious processing.

This framework motivates restricting our analysis to large-scale neural dynamics and treating cloud functions as fields over a homogeneous space. The connectome harmonics formalism also enables the use of the Schr\"odinger equation to model rare long-range interactions in the brain \citep{Haggarty2025} and underlies neural field theories that treats brain activity as fields over homogeneous spaces \citep{Atasoy2016}.

\textbf{3}. Theta and alpha oscillations forming cortical traveling waves are considered a mechanism for large-scale neural coordination \citep[][for a detailed discussion]{Zhang2018,Mohan2024}. Their phase dynamics exhibit gradual, partially synchronous spatio-temporal variations, which can reflect interactions between bottom-up and top-down processes \citep{Xu2023}. Models such as the Kuramoto and Hopf models, widely used to simulate such brain oscillations \citep[e.g.,][for a review]{NartalloKaluarachchi2026}, share a key property: \textit{invariance under global phase shifts}. This reflects the fact that only relative phase differences are functionally relevant, while global phase offsets can be absorbed by a change of variables. The corresponding governing equations may be nonlinear, as long as their nonlinear properties are invariant under such transformations. 

In our framework, we adopt this invariance as a fundamental feature of long-range neural dynamics, while possible deviations are left for future study.

\section{Dynamics of Fields with Global Phase-Shift Invariance}\label{App:C}

In this section, we consider a neural field model as a specific instance of the neural field approach to clarify its relationship with the cloud-function formalism introduced next. On short timescales, neural field theories describe collective neuronal firing via a sigmoidal transfer function with finite propagation delays \citep[][for a review]{Cook2022}. On longer timescales, however, oscillations can be approximated using polynomial nonlinearities with quasi-local delay effects, yielding forms akin to the complex Ginzburg--Landau equation \citep{Cooray2024,Cooray2025}. The present analysis focuses on this long-timescale regime.
%
%

\subsection{Connectome harmonics as a psycho--neural bridge}

As discussed in the Introduction, the spatio-temporal theory of consciousness \citep{Northoff_2022} and its extension, the theory of common currency \citep{Northoff2025b}, posit a second-order ``isomorphism,'' or a dynamical bridge, between stimulus-evoked cortical activity and perceptual content, including the \textit{embedding} of object representations in mental space-time. This embedding motivates the notion of space-time clouds---regions with blurred boundaries that capture perceptual uncertainty \citep{LubPlav2021}. Building on this framework, we treat supraliminal object representations as spatial projections (cloud functions) of these clouds. The present subsection outlines the key theoretical constructions supporting the proposed approach within visual perception.

\textit{First}, visual representations in the cortex arise from hierarchical and parallel processing, along with dynamic remapping across topographically organized visual areas \citep[e.g.,][for a review]{Lima2023}. Accordingly, neural activity encodes objects in a distributed, structured manner shaped by cortical topology and connectivity. Consequently, Wilson--Cowan-type neural field models capture both stimulus properties and network structure, reproducing perceptual phenomena such as visual illusions \cite{Bertalmio2020}, bottom-up/top-down interactions \cite{Schummers2005}, and multiple coding schemes, including large-scale oscillations and local spike-rate dynamics \cite{Nesse2024}.

To elucidate our subsequent constructions, we outline a typical Amari-type neural field model recently analyzed in detail by \citet{Cooray2024,Cooray2025}. The model describes the dynamics of two neural fields $\phi(x,t)$ and $\varphi(x,t)$, representing excitatory ($\phi$) and inhibitory ($\varphi$) population activity across the cortical sheet $\Upsilon$,  governed by the coupled equations
\begin{equation}\label{eq:WC72}
	\begin{split}
		\tau\frac{\partial \phi}{\partial t} & =  -\phi +  \widehat{\mathcal{W}}_{ee}\cdot S(\phi) - \widehat{\mathcal{W}}_{ei}\cdot S(\varphi)\,,\\
		\tau\frac{\partial \varphi}{\partial t} & =  -\varphi +  \widehat{\mathcal{W}}_{ie}\cdot S(\phi) - \widehat{\mathcal{W}}_{ii}\cdot S(\varphi)\,.
	\end{split}
\end{equation}
Here, $\tau$ denotes the characteristic temporal scale of the neural dynamics; the integral operators $\widehat{\mathcal{W}}_{\alpha\alpha'}$ act on a function $\varpi(x')$ defined over $\Upsilon$ as 
\begin{equation}\label{eq:WC72RR}
	\Big(\widehat{\mathcal{W}}_{\alpha\alpha'}\cdot \varpi \Big)_x = \int_\Upsilon \mathcal{W}_{\alpha\alpha'}(x,x') \varpi(x')\,dx'
\end{equation}
and describe long-range synaptic input from the $\alpha'$- to $\alpha$-population via the spatial kernels $\mathcal{W}_{\alpha\alpha'}(x,x')$;  and the function $S(\cdot)$ is a sigmoid nonlinearity, e.g., 
\begin{equation}\label{eq:WC72SS}
	S(u_\alpha) = \big[1 + e^{-u_\alpha} \big]^{-1} \qquad \text{($u_1 = \phi$, $u_2 = \varphi$)}
\end{equation}
representing firing responses. For large-scale neural activity patterns, this threshold-like nonlinearity can often be approximated by polynomial expansions \cite{Cooray2024,Cooray2025}.

\textit{Second}, neural field formalism has been widely applied to model cognitive phenomena, including the neural correlates of consciousness and the distinction between normal and pathological states based on EEG data \citep[][for a review]{Polyakov2025}, as well as the interplay between spike-rate and phase coding \citep[][for a review]{Nesse2024} and between stable and dynamic coding in working memory \citep[][for a review]{Stroud2024}. Dynamical field theory, an extension of neural field theory, has been developed to address higher cognitive processes \citep[][for an introduction]{Sabinasz2023} and \citep[][for details]{Schoener2016Book}.

However, neural field models typically assume a one- or two-dimensional Euclidean spatial manifold, rendering comparisons with empirical observations largely phenomenological. Direct correspondence with physical measurements---such as visual scenes or data from fMRI, EEG, and MEG---remains challenging due to the structural complexity of the visual cortex. Although evidence indicates that the cortex implements multiple topology-preserving mappings of visual space \citep{Wandell2007}, reconstructing these mappings from data (e.g., fMRI) remains an open problem. Only recently have topology-preserving methods for such data been proposed (see, e.g., \citep{Tu2021,Tu2022,Ta2022,Xiong2026}; for a review, see \citep{Ribeiro2025}).

\textit{Third}, the formalism of \textit{connectome harmonics} \citep{Atasoy2016} provides an alternative approach to cortical complexity by describing cortico-cortical interactions through the structural connectivity of the human brain---the human connectome $\mathfrak{C}$. The basic elements of this description are the eigenvectors  $\{\xi_\lambda(j)\}$ of the symmetric graph Laplacian \citep{Chung1997}, defined over the vertices $\{j\}$ of $\mathfrak{C}$  by 
\begin{equation}
\label{NN:1}
	\Delta_\mathfrak{C}\cdot\xi_\lambda = \lambda \xi_\lambda\,, \ \text{where}\ \Delta_\mathfrak{C} = \mathbf{I} - \mathbf{D}^{-1/2} \mathbf{A} \mathbf{D}^{-1/2} ,
\end{equation}
$\mathbf{A}=\|A_{ij}\|$ denotes the connectome adjacency matrix, and $\mathbf{D}=\|D_i = \sum_j A_{ij}\|$ is the diagonal matrix of node degrees. Small eigenvalues correspond to harmonics describing large-scale cortical interactions and low-frequency oscillatory activity, whereas large eigenvalues characterize short-range interactions and high-frequency components. Neural field activity can thus be expanded as
\begin{equation}\label{eq:CH}
	u_\alpha(j,t)=\sum_\lambda a_{\alpha,\lambda}(t)\xi_\lambda(j)\,,\quad u_\alpha=\phi,\varphi\,,
\end{equation}
yielding a natural multiscale decomposition of cortical dynamics that separates spatial organization from temporal evolution \citep[][for a review]{Atasoy2018Rev}.

This approach can be extended to \textit{functional harmonics} associated with the brain's functional connectivity graph \citep{Glomb2021}, where structure--function coupling enables the construction of indicators of consciousness \citep{Luppi2023}. The connectome harmonic formalism also generalizes the Wilson--Cowan neural field model by replacing the underlying Euclidean space with a graph topology \citep{Aqil2021}.


\textit{Fourth}, substantial evidence indicates that neural representations of imagined and perceived stimuli arise from largely overlapping processes involving high-level brain networks \cite[e.g., for a review][]{Dijkstra_2019,VanCaenegem2024,Anderson2026}, with unconscious mechanisms playing a central role in both domains \cite{Nanay2020}. In addition, mental imagery has been characterized as a form of \textit{reversed neural processing} relative to the feedforward hierarchy of sensory processing \cite{Dijkstra_2020}, a reversal that is also observed in perception-based memory reconstruction \cite{Linde_Domingo_2019}. Moreover, visuospatial coding extends beyond classical visual areas, suggesting a domain-general representational scaffold that links neural computation to perception--action loops \cite{Groen2022}.

\begin{figure}
	\centering
	\includegraphics[width=0.8\columnwidth]{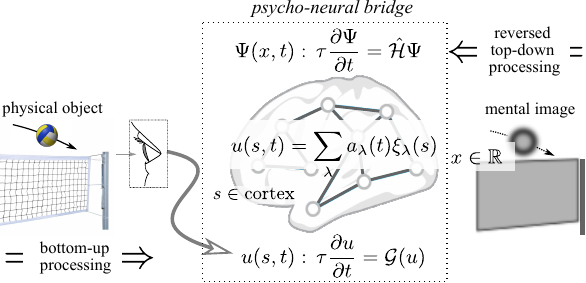}
	\caption{Schematic representation of the psycho--neural bridge based on connectome harmonics $\{\xi_\lambda(s)\}$, within which neural activity patterns are described, in the framework of neural field theory, from two complementary perspectives: bottom-up processing and reversed top-down processing. The blurred boundaries of mental images reflect uncertainty in its embedding within the mental representation of environmental space.}
	\label{F:Bridge}
\end{figure}

The four issues discussed above form the premises of our approach to describing the evolution of high-level representations of perceived stimuli. The connectome harmonics---more precisely, their amplitudes---are assumed to serve as a \textit{bridge} between bottom-up sensory processing and high-level processing underlying the formation of mental images, such as perceived objects and their environmental embedding (Fig.~\ref{F:Bridge}). Reversing this high-level process allows one to characterize the dynamics of harmonic amplitudes in terms of features of conscious content. However, unlike our previous constructions \cite{LubashNatPhen2025}, the proposed formalism treats conscious and unconscious aspects of high-level representations as integral components. Accordingly, the following two key aspects must be taken into account: 
\begin{enumerate}[topsep=0.1\baselineskip,itemsep=0.1\baselineskip, leftmargin = \parindent, label= \,\arabic*)]
	\item the oscillatory nature of neural activity, a fundamental feature of brain dynamics underlying cognitive phenomena \citep[][for a review]{SingerEffenb2025,Northoff2025a};
	\item the holistic, integrated nature of conscious images \citep{Albertazzi_2016}.
\end{enumerate}

Put differently, this psycho-neural bridge should enable the integration of the basic characteristics of neural field dynamics and the features of consciously recognized mental images within a single formalism.
 
\subsection{Neural field of supraliminal processing}

The aforementioned Aspect~1 motivates us to model supraliminal information processing using the formalism of neural field theory, which (\textit{i}) resembles model~\eqref{eq:WC72} and (\textit{ii}) incorporates the basic characteristics of the corresponding mental image, including, first, the spatial structure of the environment. In particular, the underlying manifold over which the neural fields $\{\phi, \varphi\}$ are distributed can be identified with the perceived physical space $\mathbb{R}$. Second, for example, the localization of an observed object in the environment, as perceived under physiological and attention-dependent cognitive uncertainties, should be reflected in the dynamics of the neural field. Mathematically, this is taken into account by introducing integral operators acting on the neural field components ${u_1 = \phi, u_2=\varphi}$, such as (cf. Exp.~\ref{eq:WC72RR})
\begin{equation}\label{NFL}
	\Big(\widehat{\mathcal{W}}_{\alpha\alpha'}\cdot S[u_{\alpha'}]\Big)_x = \int_\mathbb{R} \mathcal{W}_{\alpha\alpha'}\big(x,x'\mid\Omega\big) S\big[u_{\alpha'}(x')\big]dx',
\end{equation}
where the kernel $\mathcal{W}_{\alpha,\alpha'}\big(x,x'\mid\Omega\big)$ also depends on the potential landscape $\Omega(x')$ in the space of high-level representations, which arises from low-level sensory processing and reflects the structure of the environment. In the present paper, we confine our analysis to large-scale neural fields for which the sigmoid nonlinearity $S(\cdot)$ can be approximated by a polynomial nonlinearity.

Aspect~2, in the context of ``reversing'' high-level processing, motivates us to endow neural fields with an additional feature. Specifically, we assume that uncertainty in human perception---for example, when a subject evaluates the position, size, or shape of an observed object---is grounded in the uncertainty associated with encoding such characteristics within neural networks. The concept of space-time clouds \cite{ihor2017bookmind,LubPlav2021}, reflecting uncertainty in human perception and cognitive evaluation, emphasizes that the mental image of an observed object, together with its embedding, for instance, in the mental representation of the physical space-time continuum, constitutes a \textit{holistic, integrated entity}. This implies that cognitive properties of such images cannot be attributed to their individual parts; rather, they must be ascribed to the entity as a whole.\footnote{Quantum particles exemplify such holistic, integrated entities. Specifically, the physical properties of a quantum particle are attributed to its quantum state, represented by the wave function, as an indivisible whole rather than to particular spatial regions or localized fragments of the wave function. This reflects the holism and nonseparability of quantum systems \citep[e.g.,][]{sep-physics-holism}.} Therefore, the neural field representation of a space-time cloud---the cloud function $\Psi$---should not be treated merely as a particular form of neural field fragment in the space $\mathbb{R}$. Rather, it must be regarded as an entity with internal integrity.

As a consequence, the resulting equation governing the dynamics of such neural fields must operate on cloud functions as fundamental entities. Put differently, the desired model of a neural field underlying supraliminal processing has to
\begin{itemize}[topsep=0.1\baselineskip,itemsep=0.1\baselineskip, leftmargin = \parindent, label=--]
	\item admit a reduction to an equation governing the dynamics of a cloud function $\Psi$ in terms of $\Psi$ itself, possibly together with its complex conjugate $\Psi^*$ and the potential landscape $\Omega$ generated by low-level bottom-up sensory processing;
	\item preserve the integrity of the cloud function $\Psi$, reflecting the fact that, if an external object is perceived, it is observed as being in some state, possibly including the state of invisibility.
\end{itemize}
Mathematically, this means that the desired governing equation can be generally written in the form
\begin{equation}\label{NFSP:1}
	\frac{\partial \Psi}{\partial t} = \hat{\mathcal{F}}(\Psi,\Psi^*,\Omega) \quad \Rightarrow \quad \|\Psi\| = 1\,,
\end{equation}
where $\hat{\mathcal{F}}(\ldots)$ is a (possibly) nonlinear operator, and $\|\Psi\|$ denotes a norm of cloud functions reflecting their integrity. Since the cloud function $\Psi$---the basic neural representation of the observed object---is extended in the space $\mathbb{R}$, conscious comparison of two observed objects can be implemented in neural processing only via a certain overlap $\langle \Psi_1 | \Psi_2 \rangle$ of their cloud functions $\Psi_1$ and $\Psi_2$. The higher the value $\bra{\Psi_1}\ket{\Psi_2}$, the closer the objects are to each other, and we may impose the condition $\|\Psi\| = \braket{\Psi}^{1/2} = 1$ on the cloud function because any perceived object coincides with itself \cite{LubPlav2021}. The cloud-function overlap $\bra{\Psi_1}\ket{\Psi_2}$ is a bilinear form, which allows us to treat the space $\mathbb{H}$ of cloud functions as a Hilbert space.

The following two subsections develop these ideas. 

\subsection{Neural field dynamics with global phase-shift invariance}\label{Sec:3.1}

Within the proposed model, a neural field state is specified by a real-valued vector---the state vector 
\begin{equation}\label{AppC:1}
 \Phi(x,t) = \begin{Vmatrix} \phi(x,t) \\ \varphi(x,t)\end{Vmatrix},\quad x\in \mathbb{R},  
\end{equation}
whose dependence on time $t$ reflects the dynamics of neural activity. Without loss of generality,\footnotemark{} the inner product of two state vectors $\Phi_1$, $\Phi_2$, together with the induced norm, is assumed to be defined as
\begin{equation}\label{AppC:2}
  \braket{\Phi_1}{\Phi_2}_\mathbb{R} = \int_{\mathbb{R}} \big[ \phi_1(x,t) \phi_2(x,t) + \varphi_1(x,t) \varphi_2(x,t)  \big]dx\,.   
\end{equation}
Here, the time variable $t$ is treated as fixed.

\footnotetext{If the inner product is initially introduced in the Hilbert space $\mathbb{H}$ as a general positive-definite bilinear form, then, by a linear transformation $\widetilde{\Phi} = \hat{\mathcal{T}}\Phi$, this form can be reduced to \eqref{AppC:2}.}

The state vector $\Phi(x,t)$, regarded as the supraliminal representation of a sensuously perceived physical stimulus (object), is assumed to be governed by its comparison with the subliminal representation $\Omega(x,t)$ of the same stimulus, formed at the early stage of sensory information processing. More precisely, the potential landscape of the Hilbert space $\mathbb{H}$, in which the $\Phi$-dynamics unfolds, is determined by two factors. The first factor is the subliminal representation of the external stimulus, embedded in the Hilbert space $\mathbb{H}$ as the element $\Omega(x,t)$. The second factor is the degree $A$ of top-down selective attention. Combining these two factors, we specify the landscape as the function $A\Omega(x,t)$, where, in principle, the attention degree $A(x,t)$ can itself be treated as an element of the Hilbert space $\mathbb{H}$, thereby allowing selective modulation of different regions in $\mathbb{R}$. In particular, this landscape model enabled us to describe the power law of working memory within the cloud function formalism \citep{LubashNatPhen2025}.

The $\Phi$--$\Omega$ comparison may be interpreted as the essence of predictive coding reformulated in these terms \citep{LubashNatPhen2025}. Consequently, only those regions of the space $\mathbb{R}$ that carry information about the system state at the present time---namely, regions in which at least one of the two conditions $\phi(x,t)\neq 0$ or $\varphi(x,t)\neq 0$ holds---can induce temporal variations of $\Phi(x,t)$, including variations at spatially distant points through nonlocal effects. For this reason, a plausible governing equation for the state-vector dynamics must be quasi-linear.\footnote{It is necessary to remain that the proposed model is restricted to large-scale neural activity patterns.}

In the present paper, we ignore random effects in the system dynamics, which require separate investigation. Under this assumption, the governing equation for the dynamics of the state vector $\Phi(x,t)$ can be written in matrix form as:
\begin{equation}\label{AppC:3}
\tau\frac{\partial \Phi }{\partial t}= \hat{\mathbf{G}}\Phi \equiv
\begin{Vmatrix}
\hat{\mathcal{G}}_{\phi\phi} & \hat{\mathcal{G}}_{\phi\varphi} \\
\hat{\mathcal{G}}_{\varphi\phi} & \hat{\mathcal{G}}_{\varphi\varphi}
\end{Vmatrix}\cdot
\begin{Vmatrix}
\phi \\
\varphi
\end{Vmatrix},
\end{equation}
where $\tau\gtrsim  300$~ms is a time scale characterizing the duration of supraliminal information processing, and $\hat{\mathcal{G}}_{\{\cdot\}}$ are possibly nonlinear operators acting in the Hilbert space $\mathbb{H}$. Actions of these not necessarily Hermitian operators $\hat{\mathcal{G}}_{\{\cdot\}}$ on a function $\varpi(x)\in \mathbb{H}$ is specified as
\begin{equation}\label{AppC:3.ad1}
\Big[\hat{\mathcal{G}}_{\{\cdot\}} \varpi\Big](x) =\int_\mathbb{R} \mathcal{G}_{\{\cdot\}}(x,x',t) \varpi(x')dx'\,,
\end{equation}
where the operator kernel 
\begin{equation}\label{AppC:3.ad10}
\mathcal{G}_{\{\cdot\}}(x,x',t) = \mathcal{G}_{\{\cdot\}}\big[ x,x' \mid A\Omega, I(\Phi,\Phi)\big] 
\end{equation}
may, first, depend explicitly on time $t$ through the inclusion of the subliminal stimulus representation $A\Omega(x,t)$, amplified by external attention $A$, and, second, include terms representing global phase-shift invariants $I(\Phi,\Phi)$ of the state vector $\Phi(x,t)$, thereby endowing the operator $\hat{\mathcal{G}}_{\{\cdot\}}$ with nonlinear properties. In this paper, we consider second-order invariants admitting the following general representation:
\begin{gather}
\label{AppC:20.1}
  I(\Phi,\Phi) =\iint\limits_{\mathbb{R}\times \mathbb{R}} \Phi^T(x,t)\cdot \hat{\mathbf{I}}(x,x')\cdot \Phi(x',t)\,dxdx'\,,
  \\
\intertext{where the kernel matrix}
\label{AppC:20.2}
\hat{\mathbf{I}}(x,x') =  
  \begin{Vmatrix}
    {\mathcal{I}}_{\phi\phi}(x,x') & {\mathcal{I}}_{\phi\varphi}(x,x') \\
    {\mathcal{I}}_{\varphi\phi}(x,x')  & {\mathcal{I}}_{\varphi\varphi}(x,x') 
  \end{Vmatrix},
\end{gather}  
due to the integral structure of the bilinear form~\eqref{AppC:20.1}, may be assumed \textit{a priori} to be symmetric. Specifically, the following equalities  
\begin{equation} \label{App:N1}
  \mathcal{I}_{\phi\phi}(x,x') = \mathcal{I}_{\phi\phi}(x',x)\,, \quad \mathcal{I}_{\varphi\varphi}(x,x')  = \mathcal{I}_{\varphi\varphi}(x',x)\,, \quad \mathcal{I}_{\phi\varphi}(x,x') = \mathcal{I}_{\varphi\phi}(x',x)\\
\end{equation}
are assumed to hold. Here, the superscript $T$ denotes matrix transposition, and the kernels $\mathcal{I}_{\{\cdot\}}(x,x')$ define linear integral operators acting on functions over $\mathbb{R}$. Possible singularities in the spatial structure of these operators motivate the introduction of differential operators, such as the gradient and the Laplacian, acting on the state vector $\Phi(x,t)$.  For the kernels of the form 
\begin{equation}\label{KerNorm} 
\mathcal{I}_{\phi\phi}(x,x')=\mathcal{I}_{\varphi\varphi}(x,x')=\delta(x-x')\quad\text{and}\quad \mathcal{I}_{\phi\varphi}(x,x')=\mathcal{I}_{\varphi\phi}(x,x') = 0\,,
\end{equation}
functional~\eqref{AppC:20.1} coincides with the inner product~\eqref{AppC:2}.

The global phase-shift invariance of the $\Phi$-dynamics is expressed by the requirement that, under the transformation 
\begin{equation}\label{AppC:15}
  \begin{Vmatrix}
    \phi(x,t) \\ \varphi(x,t)
  \end{Vmatrix} = 
  \hat{\mathbf{T}}_\theta \cdot   
  \begin{Vmatrix}
    \widetilde{\phi}(x,t) \\ \widetilde{\varphi}(x,t)
  \end{Vmatrix}\quad \text{with} \quad
  \hat{\mathbf{T}}_\theta = 
  \begin{Vmatrix}
    \cos\theta & -\sin\theta \\
    \sin\theta & \cos\theta
  \end{Vmatrix},
\end{equation}   
the governing equation~\eqref{AppC:3} and, correspondingly, the functionals of form~\eqref{AppC:20.1} remain unchanged. Transformation~\eqref{AppC:15} can be interpreted as a counterclockwise rotation of the two-dimensional plane $\{\phi,\varphi\}$ around its origin by an angle $\theta$. Here, $\theta$ is treated as an arbitrary angle independent of the spatial position $x\in\mathbb{R}$; therefore, this transformation represents a global rotation of the field $\Phi(x,t)$ as a whole. Using the standard rules for the transformations of operators and bilinear forms induced by the state-vector transformation~\eqref{AppC:15}, we conclude that the global phase-shift invariance of the $\Phi$-dynamics holds if the following conditions
\begin{subequations}\label{AppC:GIc}
\begin{align}
\label{AppC:Gc}
    \mathcal{G}_{\phi\phi}(x,x',t) & = \mathcal{G}_{\varphi\varphi}(x,x',t) & \mathcal{G}_{\phi\varphi}(x,x',t) & = -\mathcal{G}_{\varphi\phi}(x,x',t)  
\\
\label{AppC:Ic}
    \mathcal{I}_{\phi\phi}(x,x',t) & = \mathcal{I}_{\varphi\varphi}(x,x',t) & \mathcal{I}_{\phi\varphi}(x,x',t) & = -\mathcal{I}_{\varphi\phi}(x,x',t)  
\end{align}
\end{subequations}
are satisfied for any pair $(x,x')$ and any time $t$. Moreover, by virtue of \eqref{App:N1} and \eqref{AppC:Ic}, the off-diagonal terms of the kernel matrix associated with any invariant $I(\Phi,\Phi)$ has to be antisymmetric under the exchange of arguments $x \leftrightarrow x'$, namely,
\begin{equation} \label{App:N2}
  \mathcal{I}_{\phi\varphi}(x,x') = -\mathcal{I}_{\phi\varphi}(x,x')\,,
\end{equation}
so, in particular, $\mathcal{I}_{\phi\varphi}(x,x)= \mathcal{I}_{\varphi\phi}(x,x) = 0$.

\subsection{Cloud function dynamics}\label{Sec:3.2}

The formalism of cloud function dynamics is based on the neural field dynamics developed in Sec.~\ref{Sec:3.1}. We begin by introducing the cloud function $\Psi(x,t)$, defined as 
\begin{equation}\label{AppC:5}
 \Psi(x,t) = \phi(x,t) + i\varphi(x,t)\,,
\end{equation} 
and regard the conversion from the neural field description based on the state vector $\{\phi(x,t),\varphi(x,t)\}$ to the cloud function formalism as an isomorphism
\begin{equation}\label{AppC:7}
\begin{Vmatrix}
  \phi \\ \varphi
\end{Vmatrix}   = \hat{\mathbf{M}}
\begin{Vmatrix}
  \Psi \\ \Psi^*
\end{Vmatrix}, 
\quad\text{where}\quad  \hat{\mathbf{M}} = \frac12\begin{Vmatrix}
  1 &1 \\-i & i
\end{Vmatrix}   
\ \ \text{and} \ \ 2\hat{\mathbf{M}}^\dagger \equiv 
\hat{\mathbf{M}}^{-1} = \begin{Vmatrix}
  1 & i \\1 & -i
\end{Vmatrix},
\end{equation}
where $\Psi^*$ denotes the complex conjugate of $\Psi$ and the superscript  $\dagger$ denotes the Hermitian conjugate of the corresponding operator. In these terms, the bilinear invariant~\eqref{AppC:20.1} reduces to  
\begin{gather}
\label{Q:1}
I(\Psi) \stackrel{\text{def}}{{}={}} \ev{\hat{\mathcal{I}}}{\Psi} = \iint\limits_{\mathbb{R}\times\mathbb{R}} \Psi^*(x,t) \mathcal{I}(x,x') \Psi(x',t)\, dxdx'\,,
\\
\intertext{where the operator $\hat{\mathcal{I}}$, acting in the Hilbert space $\mathbb{H}$,  is characterized by its  kernel}
\label{Q:2}
  \mathcal{I}(x,x')= \mathcal{I}_{\phi\phi}(x,x') - i \mathcal{I}_{\phi\varphi}(x,x')\,. 
\end{gather}
Due to \eqref{App:N1} and \eqref{App:N2}, the introduced operator $\hat{\mathcal{I}}$ is Hermitian. In particular, for the kernel $\mathcal{I}_z(x,x')=\delta(z-x)\,\delta(x-x')$, the invariant~\eqref{Q:1} takes the form $I_z(\Psi) = |\Psi(z,t)|^2$. 

Within the cloud function formalism, it is necessary to impose an additional normalization condition on the cloud function $\Psi$, namely,  
\begin{equation}
\label{Q:3}
\braket{\Psi} = \int\limits_{\mathbb{R}} \Psi^*(x,t) \Psi(x,t)\, dx = 1\,.
\end{equation}
In this representation, the integral form of the condition corresponds to the specific case of Exp.~\eqref{Q:1} for the kernels~\eqref{KerNorm}. The normalization condition~\eqref{Q:3} permits a probabilistic interpretation of the corresponding space--time cloud, analogous to the probabilistic interpretation of wave functions in quantum mechanics. A detailed discussion of this interpretation and its justification would require a separate, dedicated study; here, we simply adopt it.

The transformation of the governing equation~\eqref{AppC:3} should be adjusted, if necessary, so that the cloud-function normalization~\eqref{Q:3} is satisfied. This adjustment is permitted by including terms that involve only possible invariants, such as the bilinear invariants of the form~\eqref{Q:2}. In this way, \textit{first}, due to condition~\eqref{AppC:Gc}, the operator $\hat{\mathbf{G}}$ appearing in the governing equation~\eqref{AppC:3} reduces to
\begin{equation}\label{Q:4}
     \hat{\mathbf{G}} \ \mapsto\  \hat{\mathbf{G}}_\psi = \hat{\mathbf{M}}^{-1} \hat{\mathbf{G}}\hat{\mathbf{M}} = 
     \begin{Vmatrix}
      \hat{\mathcal{H}} & 0 \\ 0 & \hat{\mathcal{H}}^* 
     \end{Vmatrix},
\end{equation}
where the operator $\hat{\mathcal{H}}$, referred to as the \textit{Hamiltonian of supraliminal information processing}  or, for short, the \textit{supraliminal Hamiltonian}, is specified as
\begin{equation}\label{Q:5}
     \hat{\mathcal{H}} = \hat{\mathcal{G}}_{\phi\phi} - i \hat{\mathcal{G}}_{\phi\varphi}\,.
\end{equation} 
It should be emphasized that the diagonal form~\eqref{Q:4} of the operator $\hat{\mathbf{G}}_\psi$ justifies describing the dynamics of the cloud function $\Psi(x,t)$ in the standard manner used in quantum mechanics. Specifically, the time evolution of $\Psi$ is determined explicitly by $\Psi$ itself; its complex conjugate 
$\Psi^*$can appear only through terms of the form $\ev{\hat{V}}{\Psi}$, where $\hat{V}$ is a Hermitian operator. Moreover, the Hamiltonian $\hat{\mathcal{H}}$ may be a nonlinear, nonautonomous operator by including possible invariants, such as $|\Psi(x,t)|^2$. In the case of supraliminal information processing under consideration, the Hamiltonian $\hat{\mathcal{H}}$ is not Hermitian. Its Hermitian and anti-Hermitian components are singled out by introducing two Hermitian operators, $\hat{\mathcal{H}}'$ and $\hat{\mathcal{H}}''$, defined as  
\begin{equation}\label{Q:6}
     \hat{\mathcal{H}} = \hat{\mathcal{H}}' - i \hat{\mathcal{H}}'',
\end{equation} 
which yields the following explicit constructions of these operators and their kernels
\begin{gather}
\label{Q:7}
\hat{\mathcal{H}}' = \frac12\Big[\hat{\mathcal{H}}+\hat{\mathcal{H}}^\dagger \Big]\quad\text{and}\quad\hat{\mathcal{H}}'' = \frac{i}2\Big[\hat{\mathcal{H}}-\hat{\mathcal{H}}^\dagger \Big], 
\\
\label{Q:8}
\begin{aligned}
\mathcal{H}'(x,x') & = \frac12\Big[\mathcal{G}_{\phi\phi}(x,x') + \mathcal{G}_{\phi\phi}(x',x)\Big] - \frac{i}2 \Big[\mathcal{G}_{\phi\varphi}(x,x') - \mathcal{G}_{\phi\varphi}(x',x)\Big], \\
\mathcal{H}''(x,x') & = \frac{i}2\Big[\mathcal{G}_{\phi\phi}(x,x') - \mathcal{G}_{\phi\phi}(x',x)\Big] + \frac{1}2 \Big[\mathcal{G}_{\phi\varphi}(x,x') + \mathcal{G}_{\phi\varphi}(x',x)\Big].
\end{aligned}
\end{gather} 
The sign of the latter term in Exp.~\eqref{Q:6} is chosen to preserve the standard form of the Schr\"odinger equation in quantum mechanics when the dissipative component $\hat{\mathcal{H}}'$ of the Hamiltonian $\hat{\mathcal{H}}$ is absent, see Eq.~\eqref{Q:9}.

\textit{Second}, the governing equation~\eqref{AppC:3} for the state vector $\Phi = \{\phi,\varphi\}$ does not enforce its fixed normalization. By imposing the normalization condition~\eqref{Q:3} on the cloud function $\Psi$, we effectively modify the isomorphism~\eqref{AppC:7} through the replacement $\Psi\to \Psi \braket{\Psi}^{1/2}$. This modification manifests itself in the governing equation for the normalized cloud function $\Psi$ through the appearance of additional terms induced by the normalization. Specifically, as can be shown by direct calculation, the governing equation for $\Psi$ admits the representation
\begin{equation}\label{Q:9}
     \tau \frac{\partial \Psi}{\partial t} = \hat{\mathcal{H}} \Psi - \ev{\hat{\mathcal{H}}'}{\Psi}\Psi + i V(t)\Psi\,,
\end{equation} 
where $V(t)$ is an arbitrary real-valued function of time $t$. The arbitrariness in the choice of $V(t)$ implies that, within the developed cloud function formalism for describing supraliminal sensory information processing, all forms of the governing equation~\eqref{Q:9} are equivalent. Indeed, under the transformation
\begin{equation}\label{Q:10}
     \Psi(x,t) \ \Longrightarrow \ \exp\bigg\{\frac{i}{\tau} \int_{t_0}^t V(t')\,dt'\bigg\} \Psi(x,t)\qquad \text{($t_0$ is some instant of time)}\,,
\end{equation} 
the last term in Eq.~\eqref{Q:9} can be eliminated. This universality of the cloud function description may be interpreted as a generalization of global phase-shift invariance, allowing the phase shift to be explicitly time-dependent.

As a special case, we rewrite the governing equation~\eqref{Q:9} in the form
\begin{equation}\label{Q:11}
     \tau \frac{\partial \Psi}{\partial t} = \hat{\mathcal{H}} \Psi - \ev{\hat{\mathcal{H}}'}{\Psi}\Psi + i \ev{\hat{\mathcal{H}}''}{\Psi}\Psi = \hat{\mathcal{H}} \Psi - \ev{\hat{\mathcal{H}}}{\Psi}\Psi \,,
\end{equation} 
which explicitly demonstrates that this equation is defined up to an arbitrary shift $\hat{\mathcal{H}}\mapsto \hat{\mathcal{H}} + W(t)$, where $W(t)$ is an arbitrary complex-valued function of time. Such a shift does not require a transformation of the cloud function $\Psi$ of the form~\eqref{Q:10}. This feature allows one to interpret the supraliminal Hamiltonian $\hat{\mathcal{H}}$, or at least its components $\hat{\mathcal{H}}'$, which is responsible for dissipation in the cloud function dynamics, as a \textit{priority Hamiltonian} determining the recognition process in supraliminal information processing.

\section{Change-of-Mind Phenomenon}

In the present context, the term \textit{change of mind} refers to situations in which a subject makes a decision by choosing between two available options on the basis of accumulating evidence. However, even after an initial decision has been made and its implementation initiated, the subject may revise that decision, select the alternative option, and ultimately implement it \citep{Resulaj2009,Fleming2016chmind}. Such changes of mind challenge classical theories of decision-making, which assume evidence accumulation up to a fixed threshold, and therefore necessitate their revision.

Currently, a number of models of decision-making that accommodate changes of mind have been proposed. These include models based on post-decision evidence accumulation with a second threshold, as well as approaches incorporating uncertainty and confidence \citep[e.g.,][for a review]{Fleming2018,Stone2022}, and models emphasizing metacognitive aspects \citep{Murphy2015,Fleming2017}. In particular, confidence, choice, and reaction time---distinct aspects of evidence accumulation---may be grounded in a common neural mechanism \cite{vanderberdg2016elife}. Furthermore, changes of mind can occur even in the absence of new post-decision evidence \cite{Atiya2020}, and their properties depend not only on relative but also on absolute evidence magnitudes \citep{Turner2021,Ko_2022}.

As an alternative approach to decision-making with possible changes of mind, the concept of a graded decision process \citep{Xie2024} is also worth noting. This framework suggests that the human brain is capable of processing evidence and making choices in a flexible, analogue manner, rather than in a switch-like fashion based on reaching a fixed threshold.

\subsection{Model}

\begin{figure}
	\centering
	\includegraphics[width=0.9\columnwidth]{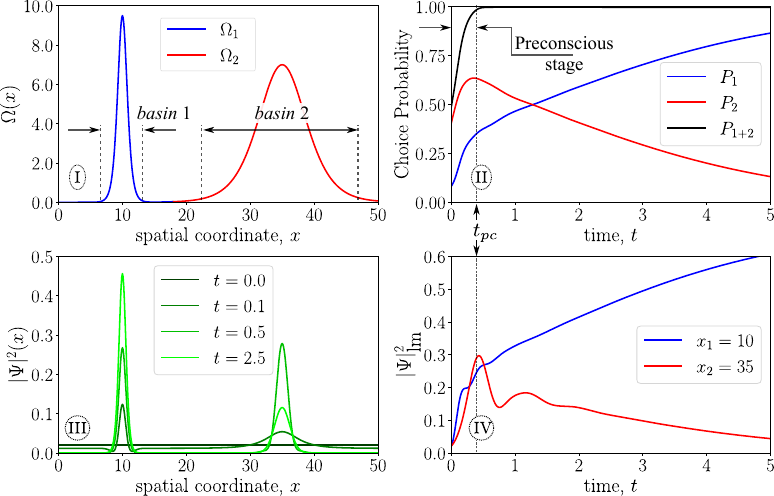}
	\caption{Characteristics of the $\Psi$-dynamics underlying the proposed mechanics of the change-of-mind phenomenon, obtained from the numerical solution of Eq.~\eqref{CM:2}. Panel~I shows the potential $\Omega(x)$, which reflects the properties of the choice options $\Omega_i$ ($i=1,2$). Panel~II presents the time dependence of the quantities $\{P_i(t)\}$ (as well as their sum $P_{12}(t)$), specified by Exp.~\eqref{CM:4} and interpreted as the probabilities of choosing the options $\{\Omega_i\}$. Panels~III and IV illustrate the evolution of the cloud-function magnitude $|\Psi(x,t)|^2$: Panel~III displays its spatial distribution at several time points, while Panel~IV shows its time dependence at the points $x_1$ and $x_2$, corresponding to the local maxima of $\Omega(x)$. The numerical simulations were performed with the following parameter values: $x_1=10$, $x_2=35$, $A_1=9.5$, $A_2=7$, $\sigma_1=1$, $\sigma_2=5$ (parameters of $\Omega(x)$, see Exp.~\eqref{Omega}), and $c_\omega = 3$, $c_\eta = 6$ (parameters of the Hamiltonian $\hat{\mathcal{H}}$, see Exp.~\eqref{CM:1}).}
	\label{F1}
\end{figure}

Adopting the cloud function formalism \citep{LubashNatPhen2025}, we assume that subliminal processing of sensory information---the early stage of information processing---gives rise to a potential landscape $\Omega(x)$ within which supraliminal information processing unfolds. The bimodal form of $\Omega(x)$ (Fig.~\ref{F1}\,I) represents the characteristics of two decision-making options, $i = 1,2$, taking into account both the properties of the external physical stimulus and the properties of the sensory organs responsible for subliminal information processing.
The difference between the two options is assumed to admit a single-parameter description, which motivates us to confine our analysis to the Hilbert space $\mathbb{H}$ defined over the one-dimensional vector space $\{x: 0\leq x\leq L\}\subset \mathbb{R}$, where $[0,L]$ denotes the decision-making domain. The potential $\Omega(x)$  is characterized by the amplitudes $\{A_i\}$, reflecting the objective preferences for option~$i=1,2$; the variables $\{x_i\}$, representing the relative difference between the two options; and the parameters $\{\sigma_i\}$, which determine the width of the basins  associated with these options. 
Below we specify $\Omega(x)$ as the inverse P\"oschl--Teller potential  
\begin{equation}\label{Omega}           
  \Omega(x) = \sum_{i=1,2} \frac{A_i}{\cosh^2[(x-x_i)/\sigma_i]}\,,
\end{equation}
whose spectral properties are well know in quantum mechanics. Here, the amplitudes $\{A_i\}$ also incorporate the degree of selective attention, which is assumed to be fixed in the present analysis.

To illustrate the proposed mechanism underlying the change-of-mind phenomenon, we may neglect essentially nonlocal effects in the dynamics of the cloud function $\Psi$. In other words, we restrict our consideration to the regime in which the action of the integral operators~\eqref{AppC:3.ad1} is primarily determined by small spatial scales. Equivalently, we assume that their kernels---and, consequently, the kernels~\eqref{Q:8} of the operators $\hat{\mathcal{H}}'$ and $\hat{\mathcal{H}}''$---possess singularities at $x' = x$, which provide the dominant contribution to the integral~\eqref{AppC:3.ad1}. Moreover, at these scales the kernels $\mathcal{G}_{\{\cdot\}}(x,x')$ may be assumed to be symmetric under the exchange $x \leftrightarrow x'$, since, for example, the observed objects do not move.

Under these assumptions, the priority Hamiltonian $\hat{\mathcal{H}}$~\eqref{Q:6} can be approximated as
\begin{equation}\label{CM:1}
	\hat{\mathcal{H}} = \kappa \Big[ (1 + i c_\omega) \Omega(x) + d^2 (1 + i c_\eta) \nabla^2 \Big],
\end{equation}
where $\nabla \equiv \partial /\partial x$ and the positive parameters $\kappa$, $d$, $c_\omega$, and $c_\eta$ reflect the integral properties of the kernel singularities.
The first term in Eq.~\eqref{CM:1} describes the evolution of the cloud function $\Psi$ toward the bimodal potential $\Omega(x)$, which mathematically represents supraliminal processing of sensory information and captures the essence of the dynamic competition between options~1 and~2 during the process of their selection. The second term accounts for the intrinsic limited capacity of supraliminal information processing to discriminate spatial locations. The constants $c_\omega$ and $c_\eta$ account for the oscillatory nature of neural activation patterns; therefore, their magnitudes must be much greater than unity.

Passing to dimensionless variables by introducing the rescaled time $t \to \kappa t/\tau$ and spatial coordinate $x\to x/d$, the governing equation~\eqref{Q:11} can be rewritten for  $0<x <L$ as 
\begin{multline}\label{CM:2}
	\frac{\partial \Psi(x,t)}{\partial t} = \bigg[ (1 + i c_\omega) \Omega(x) + (1+ i c_\eta) \nabla^2\bigg] \Psi(x,t) \\
	{} - \bigg\{\int_0^L \Big[(1 + ic_\omega) \Omega(x') |\Psi(x',t)|^2 - (1 + i c_\eta) \big|\nabla \Psi(x',t)\big|^2  \Big]dx'\bigg\}\Psi(x,t) \,,
\end{multline} 
subject to the periodic boundary conditions
\begin{equation}\label{CM:3}
	  \Psi\big|_{x=0} = \Psi\big|_{x=L}\,,\qquad \nabla \Psi\big|_{x=0} = \nabla \Psi\big|_{x=L}\,.
\end{equation} 
imposed on the cloud function $\Psi$.

\subsection{Results and discussion}

Equation~\eqref{CM:2} was solved numerically using the CN--AB2 scheme---combining the second-order Crank--Nicolson and Adams--Bashforth time-integration methods \citep[e.g.,][]{LeVeque2007}---based on the Sherman--Morrison generalization of the Thomas algorithm \citep[e.g.,][Sec.~2.7]{Press2007} and supplemented by control of the cloud function normalization. The dimensionless integration time and time step were set to $T = 10$ and $\Delta t=0.001$, respectively. The size of the computational domain and the spatial grid spacing were chosen as $L=50$ and $\Delta x = 0.01$, respectively. The remaining model parameters used in the numerical simulations are specified in the caption to Fig.~\ref{F1}. 

Panel~I of this figure shows the bimodal potential $\Omega(x)$ defined by Eq.~\eqref{Omega}, whose two peaks represent the available options $\Omega_1$ and $\Omega_2$ in the decision-making process. Option $\Omega_1$ is assumed to have higher priority than option $\Omega_2$, i.e., $A_1 > A_2$, but its basin of attraction is smaller than that of $\Omega_2$, as quantified by the spatial scales $\sigma_1 < \sigma_2$. Here, the term \textit{basin} refers to a region such that, if the interaction between the peaks were suppressed, a cloud function initially localized within the basin of option $\Omega_i$ would be ``trapped'' by that peak, i.e., it would evolve toward a small neighborhood of its maximum. In this setting, one may expect that the subject initially chooses option $\Omega_2$ and only subsequently recognizes the higher priority of option $\Omega_1$, thereby revising the initial decision. It is worth noting that we originally hypothesized that competition between two options---one of higher priority but lower ``visibility,'' and the other characterized by the opposite features---underlies the change-of-mind phenomenon.

Panels II--IV in Fig.~\ref{F1} illustrate the identified characteristics of the $\Psi$-dynamics, with particular emphasis on the quantities
\begin{equation}\label{CM:4}
	P_i(t) = \int_{\mathbb{B}_i} \big|\Psi(x,t)\big|^2 dx\,,
\end{equation}
which, for $i=1,2$, are interpreted---following~\citep{LubashNatPhen2025}---as the probabilities $P_i$ of selecting option $\Omega_i$ upon completion of the preconscious stage of supraliminal information processing, when its outcome becomes accessible to consciousness.
Here, the basin of attraction of option $\Omega_i$ is approximated by the interval $\mathbb{B}_i = (x_i - 2\sigma_i, x_i + 2\sigma_i)$. This approximation is justified by the spatial profile of the potential $\Omega(x)$ specified by Eq.~\eqref{Omega} and shown in Panel~I.
In particular, Panel~II presents the temporal evolution of the quantities ${P_i(t)}$ obtained from numerical simulations, along with their sum, $P_{12}(t) = P_1(t) + P_2(t)$. Panel~III illustrates the underlying features of the evolution of the $|\Psi(x,t)|^2$ distribution, while Panel~IV depicts in detail the dynamics of $|\Psi(x,t)|^2$ at the points $x_1$ and $x_2$, corresponding to the local maxima of $\Omega(x)$ that determine the preference for the options $\{\Omega_i\}$.

Within the proposed framework, the constructed curves support the main assumptions and hypotheses formulated above. \textit{First}, interpreting the quantities ${P_i}$ as choice probabilities is justified only if the deviation of their sum, $P_{12}$, from unity is negligible. As shown in Panel II, this condition is satisfied only after a certain time interval, $t_{pc} \approx 0.5$, following the initiation of the process. For $t \lesssim t_{pc}$, the initially homogeneous distribution $|\Psi(x,t)|_{t=0}$ evolves into two relatively narrow peaks located at the local maxima of $\Omega(x)$. These peaks can be regarded as the result of mutually independent evolution of the wave function $\Psi$ within the individual basins of attraction of the options $\Omega_1$ and $\Omega_2$, culminating in the emergence of two quasi-equilibrium states.

We interpret this phase of supraliminal information processing as a preconscious stage, the completion of which renders its output accessible to consciousness, thereby enabling the subject to consciously compare the available options. In the present work, we use a substantial decrease in the rate of evolution of $|\Psi(x,t)|^2$ as a numerical criterion for the transition from the preconscious to the conscious stage of supraliminal information processing. The theoretical description of this transition requires separate, case-specific consideration.

\textit{Second}, the subsequent evolution of $|\Psi(x,t)|^2$ illustrates the core idea of the proposed mechanism underlying the change-of-mind effect. Specifically, suppose that at time $t_{cp}$, when the output of information processing becomes accessible to consciousness, the subject selects option $\Omega_2$---whose choice probability $P_2(t_{pc})$ at that moment appears to be maximal---and begins its implementation. However, continued information processing at a slower rate subsequently reveals the higher priority of option $\Omega_1$, thereby prompting the subject to revise the initial decision and switch to implementing the option $\Omega_1$.

The non-monotonic dynamics of the cloud function $\Psi(x,t)$, illustrated in Panel~IV of Fig.~\ref{F1} for its two amplitude maxima at $x_1$ and $x_2$, is associated with the oscillatory nature of neural pattern activity governed by the operator $\hat{\mathcal{H}}''$, which constitutes a component of the priority Hamiltonian $\hat{\mathcal{H}}$.

\section{Conclusion}

In the present paper, we have developed a formalism of cloud functions that describes supraliminal processing of sensory information, encompassing both preconscious (supraliminal and unattended) and conscious (supraliminal and attended) stages.  The key aspects of the cloud function formalism were introduced in \citep{LubPlav2021,LubashNatPhen2025}. Here, we extend this framework to propose a novel mathematical description of the dynamical relationship between conscious perception of external stimuli and the underlying sensory information processing in the brain. The conceptual foundations of our approach are provided by the theory of common currency \citep{Northoff2025b}, the spatio-temporal theory of consciousness \citep{Northoff_2022}, the NCC framework \citep{Koch_2016,Koch_2018}, the formalism of connectome harmonics \citep{Atasoy2016,Atasoy2017}, the basic features of neural field theory \citep[][for a review]{Cooray2024,Cooray2025}, and the relationship between sensory perception and mental imaginary \citep[e.g.,][]{Dijkstra_2020,Anderson2026}. In some sense, the cloud function formalism may be regarded as a particular version of neural field theory combined with the integrity of mental representations of observed objects.

The concept of the cloud function reflects two key aspects. \textit{First}, the spatial structure of the cloud function $\Psi(x,t)$ inherits the organization of the mental image of a perceived stimulus, for example, an observed physical object. Consequently, general notions of physical reality---such as space, shape, and motion---can be employed to characterize this spatial structure, which corresponds to a finite spatial region with blurred boundaries, even in the case of a point-like object. In this sense, the spatial structure of the cloud function is grounded in a first-person perspective, mapped onto neural activity patterns via the neural correlates of consciousness.

\textit{Second}, the cloud function $\Psi(x,t)$ is defined at each instant in time, and its temporal evolution can be described within the framework of dynamical systems in terms of time derivatives. This property is a fundamental feature of physical processes and likewise characterizes neural pattern dynamics on the relevant time scales.

The governing equation for the cloud function $\Psi(x,t)$ has been constructed based on a general model of neural-field dynamics which, on long timescales, satisfies the condition of global-shift invariance under phase transformations of neural-pattern oscillations. Notably, both the Kuramoto model and the Hopf model of neural-pattern oscillations satisfy this condition. The resulting governing equation can be interpreted as the Schr\"odinger equation with a nonlinear, non-Hermitian Hamiltonian $\hat{\mathcal{H}}$, supplemented by an additional term---also specified by $\hat{\mathcal{H}}$---that is structurally analogous to the corresponding term in the Lotka--Volterra model. The Hamiltonian $\hat{\mathcal{H}}$ is constructed from the operators that define the neural-field dynamics of the analyzed type.

Global phase-shift invariance does not hold when a particular phase of neural-field oscillations acquires physical significance. For example, the zero phase may be associated with a high neuronal firing intensity, which becomes important at small spatial and short temporal scales. To incorporate such effects, the proposed formalism should be modified so that the equation governing the $\Psi$-dynamics also contains terms proportional to the complex conjugate $\Psi^*$, even if such terms are small. The latter follows from the fact that the developed cloud-function dynamics is degenerate with respect to the global phase of $\Psi$ (see Eq.~\eqref{Q:9}).

By way of example, the developed model has been employed to describe the change-of-mind phenomenon in decision-making. This phenomenon arises when a subject, while choosing between two available options based on accumulating evidence, may revise an initial choice during its execution. The change-of-mind phenomenon challenges the classical drift-diffusion model of decision-making, which assumes that a subject selects an option once the accumulated evidence exceeds a certain threshold.
The proposed model explains the change-of-mind phenomenon as a result of the interplay between fast preconscious information processing---making its output accessible to consciousness---and slower conscious comparison of the options. Within this framework, the initial choice is made at the end of preconscious processing and may subsequently be revised if conscious evaluation favors the alternative option.

The analyzed scenario of decision-making with changes of mind is supported by available neurophysiological data demonstrating the persistence of neural evidence accumulation after a choice has been made \citep{Murphy2015} and is consistent with the view of decision-making as a continuous multi-stage process \citep{DuranddeCuttoli2025}. It is important to emphasize that, within the developed description of changes of mind, not only the difference in option preference but also their absolute values constitute essential factors. Therefore, this formalism may also have the potential to explain the experimentally observed dependence of the change-of-mind phenomenon on the absolute intensity of perceived stimuli, namely, the ``absolute evidence magnitude'' \citep{Turner2021,Ko_2022}.

Notably, changes of mind can also arise in the absence of post-decision evidence accumulation, i.e., when external stimuli cease to be present at the moment of, or even prior to, the initial choice \citep{Atiya2020}. To account for this effect within the proposed framework, the current model must be extended. Specifically, it is necessary to incorporate self-interaction of the cloud function, a feature commonly encountered in approaches based on the Hopf model or the complex Ginzburg--Landau equation.




\end{document}